\documentclass[proceedings]{stacs}
\stacsheading{2009}{159--170}{Freiburg}
\firstpageno{159}

\usepackage{logiclmcs}

\newcommand{\ign}[1]{}

\begin{document}

\title{ Weak MSO with the Unbounding Quantifier }
\author{M. Boja\'nczyk}{Miko{\l}aj Boja\'nczyk}
\address{University of Warsaw}
\email{bojan@mimuw.edu.pl}
\urladdr{www.mimuw.edu.pl/$\mathtt{\sim}$bojan}
\thanks{Author supported by Polish
    government grant no. N206 008 32/0810.}

\keywords{automata, monadic second-order logic}

\begin{abstract}
  A new class of languages of infinite words is introduced, called the
  \emph{max-regular languages}, extending the class of
  $\omega$-regular languages. The class has two equivalent
  descriptions: in terms of automata (a type of deterministic counter
  automaton), and in terms of logic (weak monadic second-order logic
  with a bounding quantifier). Effective translations between the
  logic and automata are given.
\end{abstract}

\maketitle

\section{Introduction}
This paper introduces a new class of languages of infinite words,
which are called \emph{max-regular languages}, and include all
$\omega$-regular languages. Max-regular languages can be described in
terms of automata, and also in terms of a logic. A typical language in
the class is the property ``the distance between consecutive $b$'s is
unbounded'',
i.e.~the language 
\begin{equation}
  \label{eq:L}
  L =   \set{a^{n_1}ba^{n_2}ba^{n_3}\ldots : \forall m\  \exists i\  n_i >
    m}\ .
\end{equation}

A practical motivation can be given for considering properties that
speak of bounded distance;  e.g.~a formula of the logic in this
paper could specify that a system responds to requests with bounded
delay. We will begin, however, with a more fundamental motivation,
which is the question: what is a regular language of infinite words?

There is little doubt as to what is a regular language of finite
words. For instance, the requirement that the Myhill-Nerode
equivalence relation has finitely many equivalence classes uniquely
determines which languages of finite words should be regular. Other
notions, such as finite semigroups, or monadic-second order logic also
point to the same class.

For infinite words, however, there is more doubt. Of course, the class
of $\omega$-regular languages has much to justify calling it regular,
but some doubts remain as to its uniqueness. Consider, for instance,
the language $L$ mentioned above, or the set $K$ of ultimately
periodic words, i.e.~words of the form $wv^\omega$, say over alphabet
$a,b$.  None of these languages are $\omega$-regular. However, under
the commonly accepted definition of Myhill-Nerode equivalence for
infinite words, given by Arnold in~\cite{arnold}, both languages have exactly one
equivalence class.

Should these languages be called regular? If yes, what is the
appropriate notion of regularity? In this paper we propose a notion of
regular languages, which are called \emph{max-regular languages}, that
captures the language $L$, but not the language $K$. This new notion has
many properties that one would wish from regular languages.  The class
is (effectively) closed under boolean operations, including
negation. There is a finite index Myhill-Nerode relation, and
equivalence classes are regular languages of finite words. There is an
automaton model, there is a logical description, and translations
between the two are effective.  Emptiness is decidable. Membership is
decidable (although since we deal with infinite words, the membership
test is for certain finitely presented inputs, such as ultimately
periodic words).

So, what is this new class? One definition is in terms of logic. The
max-regular languages are the ones that can be defined by formulas of
weak monadic second-order logic extended with the unbounding
quantifier. The term ``weak'' means that only quantification over
finite sets is allowed. The unbounding quantifier $U X. \varphi(X)$ was
introduced\footnote{The quantifier introduced in~\cite{bojanboun} was
  actually the negation of $U$, saying that the size is bounded. }
in~\cite{bojanboun}, it says that the size of sets $X$ satisfying
$\varphi(X)$ is unbounded, i.e.
\begin{equation}
  \label{eq:unboundin}
  UX. \varphi(X) = \bigwedge_{ n \in \Nat}
\exists X \  \big(\ \varphi(X) \quad \land \quad  n 
\le |X| < \infty\ \big) .
\end{equation}
Monadic second-order logic with the unbounding quantifier for infinite
trees was studied in~\cite{bojanboun}, where an emptiness procedure
was presented for formulas with restricted quantification
patterns. This study was continued in~\cite{boundscolc}, where the
models where restricted from infinite trees to infinite words, but the
quantification patterns considered were more relaxed. However, no
decision procedure was given in~\cite{boundscolc} for full monadic
second-order logic with the unbounding quantifier, and the expressive
power of the logic seemed to be far too strong for the techniques used
(no undecidability results are known, though).

The basic idea in this paper is to restrict the set quantification to
finite sets (i.e.~weak quantification), while keeping the unbounding
quantifier. It turns out that with this restriction, lots of the
problems encountered in~\cite{boundscolc} are avoided, and the
resulting class is surprisingly robust. Note that for infinite words
and without unbounding quantification, weak monadic second-order logic
has the same expressive power as full monadic second-order logic; this
is no longer true when the unbounding quantifier is allowed (we prove
this using topological techniques).

The main contribution of this paper is Theorem~\ref{thm:main}, which
shows that weak monadic second-order logic with the unbounding
quantifier has the same expressive power as deterministic
max-automata. A max-automaton is a finite automaton equipped with
counters, which store natural numbers. The important thing is that the
counters are not read during the run (and therefore do not influence
the control of the automaton), which avoids the usual undecidability
problems of counter machines. The counters are only used in the
acceptance condition, which requires some counter values to be
bounded, and some to be unbounded.

To the best of the authors knowledge, quantifiers similar to the
unbounding quantifier have only been considered
in~\cite{bojanboun,boundscolc}. On the other hand, the idea to use
automata with quantitative acceptance conditions, has a long history,
going back to weighted automata of Sch\"utzenberger~\cite{weighted}
(see \cite{DBLP:journals/tcs/DrosteG07} for a recent paper on weighted
automata and related logics).

The max-automata used in this paper are closely related to an
automaton model that has been variously called a \emph{distance desert
  automaton} in~\cite{kirsten}, a
\emph{BS-automaton} in~\cite{boundscolc}, or an \emph{R-automaton}
in~\cite{krcal}.  One important application,
see~\cite{kirsten}, of these automata is
that they can be used to solve the famous star-height
problem\footnote{This is the question of calculating the least number
  of nested stars in a regular expression (without negation) that
  defines a regular language $L \subseteq \Sigma^*$. }, providing
simpler techniques and better complexities than in the famous result
of Hashiguchi~\cite{DBLP:journals/iandc/Hashiguchi88}.  (The reduction
from the star-height problem is not to emptiness of the automata, but
to something called \emph{limitedness}.)  Other problems that can be
tackled using this type of automata include the star-height of tree
languages~\cite{loedcolc} or the Mostowski index of $\omega$-regular
languages~\cite{loedcolcmostowski}.

\section{The automaton}
\label{sec:automaton}

We begin our presentation with the automaton model.

A \emph{max-automaton} has a finite set of states $Q$ and a finite set
of counters $\Gamma$. It also has a finite set of transitions. Each
transition reads an input letter, changes the state, and does a finite
sequence of counter operations. The counter operations are:

\medskip
\begin{tabular}{ll}
  $c:=c+1$.& Increment counter $c$.\\
$c:=0$. & Reset counter $c$.\\
$output(c)$. &Output the value of counter $c$.\\
$c:=max(c,d)$.& Store in counter $c$ the maximal value of counters $c,d$.
\end{tabular}
\medskip

A max-automaton is run on an infinite word $w \in \Sigma^\omega$. A
run is an infinite sequence of transitions, with the usual requirement
on consistency with the letters in the input word. Fix a run
$\rho$. With each counter $c \in C$, we associate the sequence counter
values $\rho_c \in \Nat^* \cup \Nat^\omega$ that have been output by
the instruction $output(c)$. These outputs are used by the accepting
condition, which is a boolean combination of clauses: ``the sequence
$\rho_c$ is bounded''.

Note that with this acceptance condition, it is only the set of values
in $\rho_c$ that matters, and not their order or multiplicity. This is
unlike the parity condition (where multiplicity is important), or the
S-condition of \cite{boundscolc}, where the sequence $\rho_c$ is
required to tend to infinity.

The toolkit of counter operations could be modified without affecting
the expressive power of max-automata. For instance, we could have an
operation $c:=d$, which is equivalent to $c:=0$ followed by
$c:=max(c,d)$.  On the other hand, the output instruction can be
removed (in this case, $\rho_c$ would contain all values of the
counter during the run). The output operation can be simulated by the
others as follows: for every counter $c$, we add a new output counter
$c'$, which is never incremented. Instead of doing $output(c)$, we do
$c':=c$. This way, the counter $c'$ gets only the values that were
output on the original counter $c$.

\begin{thm}\label{thm:emptiness}
  Emptiness is decidable for max-automata.
\end{thm}
\proof 

The difficulty in the proof is dealing with the max operation.

We will reduce the problem to a result from~\cite{boundscolc}. A
direct and elementary proof can also be given.  A \emph{U-automaton}
is a max-automaton that does not use the max operation, and where the
acceptance condition is a \emph{positive} boolean combination of
clauses ``counter $c$ is unbounded''.

Let $\Aa$ be a max-automaton that we want to test for emptiness.  As
is often the case, we will be searching not for an input word accepted
by $\Aa$, but for an accepting run of $\Aa$ (which is also an infinite
word).  Fix a single clause in the accepting condition, e.g.~
``counter $c$ is unbounded''. Below, we will show that the set of runs
which satisfy this clause can be recognized by a nondeterministic
U-automaton.  In particular, the set of accepting runs of $\Aa$ is a
boolean combination of languages accepted by U-automata. The result
then follows from~\cite{boundscolc}, where emptiness is shown
decidable for boolean combinations of nondeterministic
U-automata\footnote{The result in~\cite{boundscolc} is for S-automata,
  which are more powerful than U-automata. It is shown that a boolean
  combination of S-automata is equivalent to a BS-automaton, which has
  decidable emptiness.}.

Before we define the U-automaton that tests if counter $c$ is
unbounded, we introduce some auxiliary definitions.  Let $c,d$ be
counters of the automaton $\Aa$. Below we define what it means for a
finite sequence of counter operations $\rho$ to transfer $c$ to $d$,
possibly with an increment. (Formally, we are defining two ternary
relations: $T(\rho,c,d)$, for transfers, and $TI(\rho,c,d)$, for
transfers with an increment.) The idea is that after executing the
operations $\rho$, the value of counter $d$ is at least as big as the
value of counter $c$ before executing $\rho$. The definition of transfers is by
induction on the length of $\rho$:
\begin{itemize}
\item Every counter is transferred to itself by the empty sequence of
  operations, as well as the operations $c:=c+1$ and
  $output(c)$. Furthermore, $c:=c+1$ also transfers $c$ to itself with
  an increment.
\item The operation $c:=0$ transfers every counter to itself, except
  $c$.
\item The operation $c=\max(c,d)$ transfers every counter to itself,
  and also $d$ to $c$.
\item If a sequence of operations $\rho_1$ transfers $c$ to $e$, and a
  sequence of operations $\rho_2$ transfers $e$ to $d$, then their
  concatenation $\rho_1 \rho_2$ transfers $c$ to $d$. If either of the
  transfers in $\rho_1$ or $\rho_2$ does an increment, then so does
  the transfer in $\rho_1 \rho_2$.
\end{itemize}
Note that the transfer relation is regular in the following sense: for
any counters $c$ and $d$, the set of words $\rho$ that transfer
counter $c$ to $d$ is a regular language of finite words, likewise for
transfers with an increment.

Let $c$ be a counter. A finite sequence of positions $x_1< \cdots <
x_n$ in a run of $\Aa$ is called a \emph{$c$-loop} if for any $i < n$,
counter $c$ is transferred to itself with an increment by the subrun
between positions $x_i$ to $x_{i+1}$.  For a counter $d$, a \emph{$d$-trace} is a
sequence of positions $x_1 < \cdots < x_n < y$ such that for some
counter $c$, the positions $x_1 < \cdots < x_n$ are a $c$-loop, and
counter $c$ is transferred to $d$ by the subrun between positions
$x_n$ and $y$.

Equipped with these definitions, we are ready to define a
(nondeterministic) U-automaton that tests if counter $c$ is unbounded
in an input run.  The U-automaton has only one counter, and it accepts
if unbounded values are output to this counter. A run of this
automaton (which inputs a run of the automaton $\Aa$) proceeds as
follows. It uses nondeterminism to guess a $d$-trace $x_1 < \cdots <
x_n< y$, and it increments its counter at each of the positions
$x_i$. Once it sees position $y$, it outputs the counter value (which
is $n$), and resets the counter. It then finds another $d$-trace, and
again outputs its length, and so on.   It is not difficult to verify
the correctness of this construction.\qed

In this paper, we will be mainly interested in deterministic max-automata.

\section{The logic}
\label{sec:logic}
We consider an extension of weak monadic second-order logic, called
\emph{weak unbounding logic}. Recall that weak monadic second-order
logic is an extension of first-order logic that allows quantification
over finite sets (the restriction to finite sets is the reason for
the name ``weak''). In weak unbounding logic, we further add the \emph{unbounding
quantifier $UX$}, as defined in~(\ref{eq:unboundin}).

\begin{example}
  Consider the set $L$ from~(\ref{eq:L}). This language is not
  regular, but defined by the following formula of weak unbounding
  logic:
  \begin{eqnarray*}
  UX\   \forall x  \le y \le z \quad  \ x,z \in X\   \Rightarrow \ a(y)
    \land y \in X
  \end{eqnarray*}
\end{example}

The main result of this paper is that the logic and automata coincide, i.e.
\begin{thm}\label{thm:main}
  Weak unbounding logic defines exactly the same languages as
  deterministic max-automata.
\end{thm}

The more difficult direction in Theorem~\ref{thm:main} is presented in
Section~\ref{sec:logic-capt-autom}.  The easier direction, where an
automaton is simulated by the logic, can be shown by combining
standard techniques with the concepts from the proof of
Theorem~\ref{thm:emptiness}.  The key idea is that a formula of weak
unbounding logic can test if a set of positions $\set{x_1 < \cdots <
  x_n < y}$ forms a $d$-trace.  It is important that the automata are
deterministic, which allows a formula of weak logic to uniquely decode
the run that corresponds to the input word.

The formulas that are sufficient to simulate a deterministic
max-automaton are of a special type, which gives a normal form for
weak unbounding logic:
\begin{prop}
  Each formula of weak unbounding logic is equivalent to a boolean
  combination of formulas $UX \varphi(X)$, where $\varphi(X)$ does not
  use the unbounding quantifier.
\end{prop}
\proof
By
translating a formula into an automaton and then back into a formula.
\qed

\section{Weak bounding logic is captured by deterministic max-automata}
\label{sec:logic-capt-autom}

We now turn to the more difficult part of Theorem~\ref{thm:main},
namely showing that for every formula of weak unbounding logic there
is an equivalent deterministic max-automaton.

The proof is by induction on the size of the formula. To simplify the
proof, we use the usual technique of removing first-order
quantification, as in~\cite{thomas}. That is, first-order
quantification is replaced by three new predicates, all of which can
be recognized by the deterministic max-automata: ``set $X$ has one element'', ``set
$X$ is included in set $Y$" and ``all elements of set $X$ are before
all elements of set $Y$''. Together with weak second-order
quantification, these new three predicates can be used to simulate
first-order quantification, so the logic is the same. However, since
we have removed first-order quantification, in the translation to
automata we only have to deal with quantification over finite sets
(weak second-order quantification) and the new quantifier.

For purposes of the induction, we generalize the statement to formulas
with free variables. What is the word language corresponding to a
formula $\varphi(X_1,\ldots,X_n)$? This language contains words
annotated with valuations for the free set variables. We use the usual
encoding, where the label of a word position $x \in \Nat$ is extended
with a bit vector in $\set{0,1}^n$ that says which of the sets
$X_1,\ldots,X_n$ contain position $x$. More formally, for sets of word
positions $X_1,\ldots,X_n \subseteq \Nat$ and an infinite word $w \in
\Sigma^\omega$, we define the word
\begin{eqnarray*}
  w[X_1,\ldots,X_n]  \ \in \ (\Sigma \times
\set{0,1}^n)^\omega
\end{eqnarray*}
as follows. On position $x$, the new word has a tuple
$(a,b_1,\ldots,b_n)$, with $a$ the label of the $x$-th position of the
original word $w$, and the value of bit $b_i$ being $1$ if and only if
position $x$ belongs to the set $X_i$, for $i=1,\ldots,n$.  With this
notation, we can define the set of words satisfying a formula
$\varphi(X_1,\ldots,X_n)$ to be
\begin{eqnarray*}
L_\varphi =  \set{w[X_1,\ldots,X_n] :  w, X_1,\ldots,X_n \models
  \varphi}\ .
\end{eqnarray*}

Equipped with the above definition, we can use induction to show that the
logic is captured by automata, as stated in the proposition below. This result is the main ingredient in
the proof of Theorem~\ref{thm:main}.
\begin{prop}\label{prop:inductive}
  For every formula $\varphi$ of weak unbounding logic, the set
  $L_\varphi$ is recognized by a deterministic max-automaton.
\end{prop}

The proof is by induction on the size of the formula $\varphi$.
The induction base, which corresponds to the predicates ``set $X$ has one element'', ``set $X$ is included in
set $Y$" and ``all elements of set $X$ are before all elements of set
$Y$'' is easy, since all of these are $\omega$-regular languages, and
we have:
\begin{lem}\label{lemma:capture-omega-regular}
  Deterministic max-automata capture all $\omega$-regular languages.
\end{lem}
\proof By simulating a deterministic automaton with the Muller or
parity condition. We add a new counter $c_q$ for each state $q$ of the
automaton. Each time state $q$ appears, counter $c_q$ is incremented
and output. The counters are never reset. In a run of this automaton,
a state appears infinitely often if and only if its counter is
unbounded. Therefore, the Muller acceptance condition can be encoded
in the unbounding condition of a max-automaton.  \qed

The induction step for boolean operations---including negation---is no
more difficult, since the automata are deterministic and the accepting
condition is closed under boolean operations.  We are left with weak
second-order quantification and the unbounding quantifier.  We first
deal with weak quantification, in Section~\ref{sec:weak-exist-quant},
while the unbounded quantifier is treated in
Section~\ref{sec:bound-quant}.

\subsection{Weak existential quantification}
\label{sec:weak-exist-quant}

This section is devoted to showing:
\begin{prop}\label{prop:weak-closed}
  Languages recognized by deterministic max-automata are closed under
  weak quantification. In other words, if $L$ is a language over
  $\Sigma \times \set{0,1}$ recognized by a deterministic
  max-automaton, then there is a deterministic max-automaton recognizing
  \begin{eqnarray*}
    \set{ w \in \Sigma^\omega : w[X] \in L \mbox{ for some finite set
        $X$}}\ .
  \end{eqnarray*}
\end{prop}

A convenient way to prove this result would be to use nondeterministic
automata. Unfortunately, as we will later show, adding nondeterminism
to max-automata gives power beyond that of weak unbounding logic, so we
cannot use this strategy. We will have to do the existential
quantification directly in the deterministic automata.

The proof technique is actually very generic. It would work for any
model of deterministic automata that all $\omega$-regular languages
and satisfies some relaxed assumptions, mainly that the acceptance
condition is prefix-independent.

Fix a deterministic max-automaton $\Aa$ that recognizes $L$, with
state space $Q$.

A \emph{partial run} in an infinite word $w$ is a run that begins in
any position of the word (not necessarily the first position) and in
any state (not necessarily the initial one). In other words, this is a
word in $\bot^* \delta^\omega \cup \bot^\omega$, where $\delta$ is the
set of transitions of $\Aa$, that is consistent with the word $w$ on
those positions where it is defined (i.e.~where it is not
$\bot$). Since the automaton is deterministic, a partial run is
uniquely specified by giving the first configuration where it is
defined, this is called the \emph{seed configuration}. (There is also
the undefined partial run $\bot^\omega$, which has no seed
configuration.) Here, a configuration is a pair $(q,x)$, where $q$ is
a state and $x$ is a word position. Note that we do not include the
counter values in the seed configuration, since the acceptance
condition is not sensitive to finite perturbations.

We say that two partial runs \emph{converge} if they agree from some
position on. Equivalently, they converge if they share some
configuration, or both are undefined.  We say a set of partial runs
\emph{spans} a word $w$ if every partial run over $w$ converges with
some run from the set. Usually, we will be interested in finite sets
of spanning runs.

\begin{lem}\label{lemma:spanning-runs}
  For every word $w$, there is a set of at most $|Q|$  spanning runs.
\end{lem}
\proof We begin with some arbitrary configuration, and take the
partial run $\rho_1$ that begins in that configuration. If $\set
{\rho_1}$ is spanning, then we are done. Otherwise, we take some
partial run $\rho_2$ that does not converge with $\rho_1$, and see if
the set $\set{\rho_1,\rho_2}$ is spanning. If it is not, we add a
third partial run $\rho_3$, and so on. This process terminates after
at most $Q$ steps, because if two partial runs do not converge, then
they must use different states on each position where they are both
defined. So $|Q|$ partial runs that do not converge will use up all
the states.  \qed

To prove Proposition~\ref{prop:weak-closed}, we use a result stronger
than Lemma~\ref{lemma:spanning-runs}. We will show that not only the
spanning set of runs exists, but it can also be computed by a
(deterministic, letter-to-letter) transducer. By \emph{transducer} we
mean a finite deterministic automaton where each transition is
equipped with an output letter, from an output
alphabet~$\Gamma$. Therefore, the transducer defines a function $f :
\Sigma^\omega \to \Gamma^\omega$. The transducer does not have any
accepting conditions (using bounds or even parity or Muller), it just
scans the word and produces its output. It is easy to see that
deterministic max-automata are closed under preimages of transducers, as shown in
the following lemma.
\begin{lem}\label{lemma:preimage}
  If $f$ is a transducer and $\Aa$ is a deterministic max-automaton, then there is a
  deterministic max-automaton recognizing the set of words $w$ such that $f(w)$ is
  accepted by~$\Aa$.
\end{lem}

We now describe how the spanning partial runs will be encoded in the
output of the transducer.  When speaking of spanning partial runs, we
mean spanning partial runs of the automaton $\Aa$ in
Proposition~\ref{prop:weak-closed}. A single partial run can be
encoded as an infinite word over the alphabet $Q \times
\set{0,1}$. The idea is that $\set{0,1}$ is used as a marker, with $0$
meaning ``ignore the prefix until this position'', and $1$ meaning
``do not ignore''. Formally, an infinite word
\begin{eqnarray*}
  (q_1,a_1) (q_2,a_2), \ldots  \quad \in \quad (Q \times \set{0,1})^\omega
\end{eqnarray*}
is interpreted as the partial run which on position $i$ has $\bot$ if
$a_j = 0$ for some $j \ge i$, otherwise it has $q_i$. Note that if the
word above has infinitely many positions $j$ with $a_j=0$, then the
partial run is nowhere defined, i.e.~it is $\bot^\infty$.  If we want
to encode $n$ partial runs, we use $n$ parallel word sequences,
encoded as a single sequence over the product alphabet
 \begin{eqnarray*}
   (Q \times \set{0,1})^n\ .
 \end{eqnarray*}
 With the encoding of spanning runs defined, we are now ready to
 present the stronger version of Lemma~\ref{lemma:spanning-runs}.

 \begin{lem}\label{lem:spanning-transducer}
   Let $n=|Q|$. There is a transducer
   \begin{eqnarray*}
     f : \quad  \Sigma^\omega \quad \to \quad    ((Q \times \set{0,1})^n)^\omega
   \end{eqnarray*}
 such that for any word $w$, the output $f(w)$
   encodes  $n$ spanning partial runs.
 \end{lem}
 \proof The idea is to implement the proof of
 Lemma~\ref{lemma:spanning-runs} in a transducer.  The states of the
 transducer will be permutations of the state space, i.e.~tuples from
 $Q^n$ where each state appears exactly once. The initial state is any
 arbitrarily chosen permutation. When reading an input letter $a$ in
 state $\pi=(q_1,\ldots,q_n)$, the transducer does the following
 operations. First, it transforms each state in $\pi$ according to the
 letter $a$, giving a tuple $x=(q_1a,\ldots,q_na)$. This tuple is not
 necessarily a permutation, i.e.~there are may be some coordinates $i
 \in \set{1,\ldots,n}$ such that the state $q_ia$ appears already in
 $\set{q_1a,\ldots,q_{i-1}a}$. Let $I=\set{i_1,\ldots,i_k}$ be these
 coordinates, and let $\set{p_1,\ldots,p_m}$ be the states that do not
 appear in the new tuple $x$. These two sets have the same size,
 i.e.~$k=m$.  We can now correct $x$ to be a permutation $\sigma$, by
 replacing its coordinate $i_1$ with the state $p_1$, the coordinate
 $i_2$ with state $p_2$, and so on. Note that on a the coordinates
 from $I$, the new permutation $\sigma$ has a value unrelated to the
 one from $\pi$ (i.e.~$\sigma$ begins a new run), while on coordinates
 from outside $I$, the new permutation $\sigma$ simply continues the
 runs from $\pi$. This is signified in the output of the transducer,
 which is decorates each coordinate $i$ of the permutation $\sigma$
 with a bit, which is $0$ when $i \in I$ and $1$ otherwise. \qed

 We are now ready to prove Proposition~\ref{prop:weak-closed}. By
 properties of spanning sets of runs, a word $w \in \Sigma^\omega$
 belongs to the language of the proposition if and only if there is some
 $i=1,\ldots,n$ such that the following two properties hold:
\begin{itemize}
\item[(A)]  The $i$-th run encoded by $f(w)$ is defined (i.e.~the
  encoding does not contain infinitely many cancelling $0$s) and
  satisfies the accepting condition in the automaton $\Aa$.
\item[(B)] There is some finite set $X \subseteq \Nat$ such that the run of
  $\Aa$ over $w[X]$ converges with the $i$-th run encoded by $f(w)$.
\end{itemize}
Since deterministic max-automata are closed under union, it suffices
to show that for each fixed $i$, both properties (A) and (B) are
recognized by deterministic max-automata. For property (A), we use
Lemma~\ref{lemma:preimage} on preimages. Property (B), on the other
hand, is an $\omega$-regular property, which can be recognized by a
deterministic max-automaton thanks to Lemma~\ref{lemma:capture-omega-regular}.

\ign{A Corollary of Preposition~\ref{prop:weak-closed} is given below.
\begin{cor}\label{lemma:suffixes}
  If $L \subseteq \Sigma^\omega$ is  recognized by a
  deterministic max-automaton, then so $\Sigma^*L$.
\end{cor}}

\subsection{Unbounding  quantification}
\label{sec:bound-quant}

We now turn to the more difficult part of
Proposition~\ref{prop:inductive}, namely that deterministic max-automata are closed
under unbounding quantification.
\begin{prop}\label{prop:weak-closed-undounding}
  Languages recognized by deterministic max-automata are closed under unbounding
  quantification. In other words, if $L$ is a language over $\Sigma
  \times \set{0,1}$ recognized by a deterministic max-automaton, then
  so is
  \begin{eqnarray*}
    UL = \set{ w \in \Sigma^\omega : w[X] \in L \mbox{ for 
        arbitrarily large finite sets
        $X$}}\ .
  \end{eqnarray*}
\end{prop}

Fix a deterministic max-automaton $\Aa$ recognizing the language $L$
in the proposition.  Given a finite prefix $w \in \Sigma^*$ and a
state $q$ of $\Aa$, let $max(q,w)$ be the maximal size of a set $X$ of
positions in $w$ such that the automaton $\Aa$ reaches state $q$ after
reading $w[X]$. We claim that the sets $max(q,w)$ can be computed in
the counters of a deterministic max-automaton (not surprisingly, using
the max operation).

  \begin{lemma}
    There is a deterministic max-automaton with counters $\set{c_q}_{q \in Q}$ such that
  the value of $c_q$ after reading a prefix $a_1 \cdots a_n$ of the
  input is exactly $max(q,a_1 \cdots a_n)$.
  \end{lemma}
\ign{  \proof Let $Q$ be the state space of $\Aa$.   The numbers $max(q,w)$
  can be calculated by using max and increment:
  \begin{equation}
    \label{eq:maximum}
    max(q,a_1 \cdots a_{n}) \quad = \quad     \max_{ 
     \scriptstyle{(p,i) \in in(q,a_{n})}} max(p,a_1 \cdots a_{n-1})+i \ .
  \end{equation}
  In the above, $in(q,a_{n-1})$ denotes be the set of pairs $(p,i) \in
  Q \times \set{0,1}$ such that automaton $\Aa$ changes state from $p$
  to $q$ when reading the input label $(a_{n},i)$. We assume that
  the maximum is zero when the set $in(q,a_{n})$ is empty.  To
  update the counters $\set{c_q}_{q \in Q}$, several bookkeeping
  counters are used to calculate the appropriate subexpressions
  of~(\ref{eq:maximum}).

  \qed }

  We will use the values from the above lemma to capture the
  unbounding quantifier. However, some more effort is needed: it is
  not the case that an input word $w=a_1 a_2 \cdots$ belongs to $UL$
  if and only if the values $max(q,a_1\cdots a_n)$ are unbounded. In
  general, only the left to right implication holds. The right to left
  implication may fail since a value $max(q,a_1 \cdots a_n)$ is
  relevant only if the run of $\Aa$ over $w$ that begins in
  configuration $(q,n)$ can be extended to an accepting one over the
  rest of the word. The correct characterization is given below:

  \begin{lem}\label{lem:ul-char}
    A word $a_1 a_2 \cdots \in \Sigma^\omega $ belongs to $UL$ if and
    only if for some state $q$, the following values are unbounded
  \begin{eqnarray*}
    \set{max(q,a_1 \cdots a_n) : a_{n+1}
      a_{n+2} \cdots [\emptyset] \in (\Sigma \times \set{0,1})^\omega
      \mbox{ is accepted by $\Aa$ when starting in $q$}}
  \end{eqnarray*}    
  \end{lem}

  As suggested by the above lemma, to recognize the language $UL$ it
  would be convenient to have an extension of max-automata, where the
  automaton would have the ability to output $max(q,a_1 \cdots a_n)$
  only in case a certain property was satisfied by the suffix $a_{n+1}
  a_{n+2} \cdots $.  Below, we introduce such an extension of
  max-automata, which we call a {guarded max-automaton}. We then show
  that this extension can be simulated by a standard max-automaton,
  thus completing the proof of
  Proposition~\ref{prop:weak-closed-undounding}.

  An \emph{guarded max-automaton} is like a max-automaton, except that
  it is also allowed to use the following counter operation:

\medskip
\begin{tabular}{ll}
$if\ L\ then\ output(c)$. &Output the value of counter $c$, but only
if  the suffix of the \\ & input beginning at
  the next position belongs to $L \subseteq \Sigma^\omega$
.\\
\end{tabular}
\medskip

In the above operation, the language $L$---called the \emph{guard} of
the transition---must be a language recognized by a max-automaton
(without guards, although allowing guards would give the same result).
This new operation is all we need to recognize the language $UL$:
\begin{lem}
  If a language $L$ is recognized by a deterministic max-automaton,
  then $UL$ is recognized by a deterministic guarded max-automaton.
\end{lem}
\ign{\proof A deterministic guarded max-automaton can recognize if the
values in the set from Lemma~\ref{lem:ul-char} are unbounded. This is
because the condition `` $a_{n+1} a_{n+2} \cdots [\emptyset]$ is
accepted by $\Aa$ when starting in $q$'' is a property of the suffix
$a_{n+1} a_{n+2} \cdots$ recognized by a deterministic max-automaton.
\qed}

We will show that guarded outputs are redundant, and can be simulated
by non-guarded outputs.  This completes the proof
Proposition~\ref{prop:weak-closed-undounding}. The difficulty in the
proof below is that we are dealing with deterministic automata, while
a guard looks to the future.

\begin{prop}\label{prop:guarded-remove}
  For every deterministic guarded  max-automaton there is an
  equivalent deterministic max-automaton.
\end{prop}
\proof Let $\Aa$ be a deterministic guarded max-automaton. To simplify
notation, we assume that only one guarded operation,
\begin{eqnarray*}
o \quad =  \quad  if\ L\ then\ output(c)\ ,
\end{eqnarray*}
is used. The general case is done the same way. Let $\Bb$ be a
deterministic max-automaton recognizing the guard language $L$.

In the construction, we will use a concept of \emph{thread}. A thread
consists of a state of the automaton $\Bb$, as well as a number, which
corresponds to the value of counter $c$ output by the guarded
operation $o$. Note that a thread does not contain information about
values of the counters of automaton $\Bb$. The idea is that threads
will be alive for only finitely many steps, so the counters of $\Bb$
are not relevant. We will denote threads by $\tau$.  If $a \in \Sigma$
is an input letter, then we write $\tau a$ for the thread obtained
from $\tau$ by updating the state according to $a$ (and leaving the
number unchanged).

The (non-guarded) max-automaton $\Cc$ that simulates $\Aa$ works as
follows. At each point, the simulating automaton contains a finite set
$\set{\tau_1,\ldots,\tau_i}$ of \emph{active} threads.  There will be
at most one thread per state of $\Bb$, so the set of threads can be
stored using finitely many counters and the finite memory of the
automaton. This set of active threads is initially empty. Whenever
$\Aa$ does the guarded output operation $o$, a new active thread is
created, with the initial state of $\Bb$, and the number set to the
value of counter $c$. Furthermore, after reading an input letter $a
\in \Sigma$, the set of active threads is updated to
$\set{\tau_1a,\ldots,\tau_ia}$. If two active threads have the same
state, then they are merged, and only the greater number is kept
(using the max operation).

Similarly to the proof of Proposition~\ref{prop:weak-closed}, the
automaton $\Cc$ will also read the output of a transducer $f$ that
computes spanning partial runs of the automaton $\Bb$ used for the
guards.  Recall that the transducer $f$ outputs $n$ spanning partial
runs of the automaton $\Bb$, where $n$ is the number of states in
$\Bb$.  

The automaton $\Cc$ accepts a word $w$ if and only if there is some
$i=1,\ldots,n$ such that:
\begin{itemize}
\item[(A)] The $i$-th run encoded by $f(w)$ is defined (i.e.~the
  encoding does not contain infinitely many cancelling $0$s) and
  satisfies the accepting condition in the automaton $\Bb$.
\item[(B)] For every $m$, some thread storing a number greater than
  $m$ converges with $i$-th run encoded by $f(w)$.
\end{itemize}

Since deterministic max-automata are closed under finite union, we only need to show
the construction for some fixed $i$. As in the previous section,
property (A) is recognized by a deterministic max-automaton. For property (B), it
suffices to output the number stored in a thread $\tau$ whenever its
state is the same as in $\rho_i$. The automaton then accepts if the
numbers thus produced are unbounded.

\qed

\vspace{-0.5cm}

\section{Problems with nondeterminism}
In this section we show that nondeterministic max-automata are more
expressive than deterministic ones.

\begin{thm}\label{thm:separation}
  Nondeterministic max-automata recognize strictly more languages than
  deterministic ones.
\end{thm}
Contrast this result with the situation for Muller or parity automata,
which are equally expressive in the deterministic and nondeterministic
variants.  Since full monadic second-order can capture
nondeterministic automata by existentially quantifying over infinite
sets, the above theorem immediately implies:
\begin{cor}
  Full monadic second-order logic with the unbounding quantifier is
  stronger than weak monadic second-order with the unbounding
  quantifier.
\end{cor}

The separating language in Theorem~\ref{thm:separation} is 
    \begin{equation}
      \label{eq:separating}
          L = \set{a^{n_1}b a^{n_2}b a^{n_3}b  \ldots : \mbox{ some number
        appears infinitely often in $n_1,n_2,\ldots$}}\ .
    \end{equation}
   This language is captured by a nondeterministic max-automaton. The
   automaton uses nondeterminism to output a subsequence of
   $n_1,n_2,\ldots$ and accepts if this subsequence is
   bounded. Clearly, if it is bounded, then it contains an infinite
   constant subsequence.

   It remains to show that the language $L$ cannot be recognized by a
   deterministic max-automaton. For this, we will use topological
   complexity. In Lemmas~\ref{lemma:in-sigma-2}
   and~\ref{lemma:not-in-sigma-2}, we will show that every language
   recognized by a deterministic max-automaton is a boolean
   combination of sets on level $\Sigma_2$ in the Borel hierarchy,
   while the language $L$ is not.

   Below we briefly describe the Borel hierarchy, a way of measuring the
   complexity of a subset of a topological space. The topology that we
   use on words is that of the Cantor space, as described below. A set
   of infinite words (over a given alphabet $\Sigma$) is called \emph{open}
   if it is a union
   \begin{eqnarray*}
     \bigcup_{i \in I} w_i \Sigma^\omega \qquad w_i \in \Sigma^*\ ,
   \end{eqnarray*}
   with the index set $I$ being possibly infinite. In other words,
   membership of a word $w$ in an open set is assured already by a
   finite prefix of $w$.  For the Borel hierarchy, as far as
   max-automata are concerned, we will only be interested in the first
   two levels $\Sigma_1,\Pi_1,\Sigma_2,\Pi_2$. The open subsets are
   called $\Sigma_1$, the complements of these (the closed subsets)
   are called $\Pi_1$. Countable intersections of open subsets are
   called $\Pi_2$, the complements of these (countable unions of
   closed subsets) are called $\Sigma_2$.

   \begin{lem}\label{lemma:in-sigma-2}
     Any language accepted by a deterministic max-automaton is a
     boolean combination of $\Sigma_2$ sets.
   \end{lem}
\proof
   Fix a max-automaton $\Aa$, and a counter $c$ of this automaton. We
   will  examine the topological complexity of the set of runs of
   this automaton (here, a run is an infinite sequence of
   transitions).  For any fixed $n$, the following set of runs is
   clearly open:
\begin{quote}
  A value of at least $n$ is output at least once on counter $c$.
\end{quote}
In particular, its complement 
\begin{quote}
  All values of counter $c$ are at most $n$.
\end{quote}
is a closed set of runs. By taking a countable union of the above over
$n \in \Nat$, we deduce that the property
\begin{quote}
  The values of counter $c$ are bounded.
\end{quote}
is a $\Sigma_2$ property. In particular, the set of accepting runs of
any max-automaton is a boolean combination of $\Sigma_2$ sets. Since
the automata are deterministic, the function that maps an input word
to its run is continuous, i.e.~preimages of open sets are also
open. Since preimages of continuous functions preserve the levels of
the hierarchy, we conclude that any language accepted by a
deterministic max-automaton is a boolean combination of $\Sigma_2$
sets.
\qed

\begin{lem}\label{lemma:not-in-sigma-2}
  The language $L$ is not a boolean combination of $\Sigma_2$ sets.
\end{lem}
\proof
  Consider the mapping from $\Nat^*$ to $\set{a,b}^*\omega$ defined by
  \begin{eqnarray*}
   n_1,n_2,\ldots     \ldots \qquad \mapsto \qquad
     a^{n_1}b a^{n_2}b a^{n_3}b \ldots
  \end{eqnarray*}
  This is a continuous mapping. The language $L$ is the image,
  under this mapping, of the set $X$ of sequences in $\Nat^\omega$
  that have a bounded subsequence. The set $X$ is known not to be a
  boolean combination of $\Sigma_2$ sets, see~Excercise 23.2 in \cite{kechris}.
\qed

\section{Conclusion}
This paper is intended as a proof of concept. The concept is that
$\omega$-regular languages can be extended in various ways, while
still preserving good closure properties and decidability.  The class
presented in this paper, max-regular languages, is closed under
boolean operations, inverse morphisms, and quotients. It is not closed
under morphic images (which corresponds to nondeterminism on the
automaton side).

Some questions on max-automata are left unresolved.  Is the max
operation necessary in the automaton? In our construction, we use the
max twice: when defining the values $max(q,a_1\cdots a_n)$, and in
Proposition~\ref{prop:guarded-remove}. While in the first case, the
max operation can be avoided by a subtle use of factorization
forests~\cite{simon}, it is not clear how to show
Proposition~\ref{prop:guarded-remove} without using the max
operation. Another question is the exact complexity of emptiness. It
would be nice to get matching upper and lower bounds, even more so if
the lower bound would use acceptance conditions in DNF.

There are several other possibilities of future work. One is to
investigate weak bounding logic for infinite trees (note that we will
not capture all regular languages of infinite trees in this
case). Another possibility would be to
investigate full monadic-second order logic, or possibly other
quantifiers that can be added to weak monadic second-order logics.
 The
techniques used in this paper are fairly generic, so it seems
plausible that such quantifiers can be found.


\begin{thebibliography}{10}

\bibitem{krcal}
P.~A. Abdulla, P.~Krc{\'a}l, and W.~Yi.
\newblock R-automata.
\newblock In {\em CONCUR}, pages 67--81, 2008.

\bibitem{arnold}
A.~Arnold.
\newblock A syntactic congruence for rational omega-language.
\newblock {\em Theor. Comput. Sci.}, 39:333--335, 1985.

\bibitem{bojanboun}
M.~Boja\'nczyk.
\newblock A bounding quantifier.
\newblock In {\em Computer Science Logic}, volume 3210 of {\em Lecture Notes in
  Computer Science}, pages 41--55, 2004.

\bibitem{boundscolc}
M.~Boja\'nczyk and T.~Colcombet.
\newblock Omega-regular expressions with bounds.
\newblock In {\em Logic in Computer Science}, pages 285--296, 2006.

\bibitem{loedcolc}
T.~Colcombet and C.~L\"oding.
\newblock The nesting-depth of disjunctive mu-calculus for tree languages and
  the limitedness problem.
\newblock In {\em Computer Science Logic}, volume 5213 of {\em Lecture Notes in
  Computer Science}, 2008.

\bibitem{loedcolcmostowski}
T.~Colcombet and C.~L\"oding.
\newblock The non-deterministic mostowski hierarchy and distance-parity
  automata.
\newblock In {\em International Colloquium on Automata, Languages and
  Programming}, volume 5126 of {\em Lecture Notes in Computer Science}, pages
  398--409, 2008.

\bibitem{DBLP:journals/tcs/DrosteG07}
M.~Droste and P.~Gastin.
\newblock Weighted automata and weighted logics.
\newblock {\em Theor. Comput. Sci.}, 380(1-2):69--86, 2007.

\bibitem{DBLP:journals/iandc/Hashiguchi88}
K.~Hashiguchi.
\newblock Algorithms for determining relative star height and star height.
\newblock {\em Inf. Comput.}, 78(2):124--169, 1988.

\bibitem{kechris}
A.~S. Kechris.
\newblock {\em Classical Descriptive Set Theory}, volume 156 of {\em Graduate
  Texts in Mathematics}.
\newblock Springer, 1995.

\bibitem{kirsten}
D.~Kirsten.
\newblock Distance desert automata and the star height problem.
\newblock {\em Theoretical Informatics and Applications}, 39(3):455--511, 2005.

\bibitem{weighted}
M.~P. Sch\"utzenberger.
\newblock On the definition of a family of automata.
\newblock {\em Information and Control}, 4:245--270, 1961.

\bibitem{simon}
I.~Simon.
\newblock Factorization forests of finite height.
\newblock {\em Theoretical Computer Science}, 72:65--94, 1990.

\bibitem{thomas}
W.~Thomas.
\newblock Languages, automata, and logic.
\newblock In G.~Rozenberg and A.~Salomaa, editors, {\em Handbook of Formal
  Language Theory}, volume III, pages 389--455. Springer, 1997.

\end{thebibliography}
\end{document}